\documentclass[11pt]{paper}
\usepackage{amssymb,latexsym, amsmath}
\usepackage{amscd}

\usepackage[dvips]{graphicx}

 \newtheorem{thm}{Theorem}[section]
 
 \newtheorem{lem}{Lemma}[section]
 \newtheorem{cor}{Corollary}[section]

\newtheorem{dfn}{Definition}[section]

\numberwithin{equation}{section}

\renewcommand{\setminus}{\smallsetminus}

\def\ji {\char'032}

\def\m  {\char'176}

\font\rrm=wncyr10%
\font\rit=wncyi10

\newcommand{\A}{A^{-1}}
\newcommand{\R}{\mathbb{R}}
\newcommand{\N}{\mathbb{N}}
\newcommand{\Z}{\mathbb{Z}}
\newcommand{\p}{\partial}
\newcommand{\ds}{\displaystyle}

\DeclareMathOperator{\diag}{\mathrm{diag}}

\title{GEODESIC AND BILLIARD FLOWS ON QUADRICS IN PSEUDO--EUCLIDEAN
SPACES: L--A PAIRS AND CHASLES THEOREM}

\author{ Bo\v zidar Jovanovi\'c \and
Vladimir Jovanovi\'c}

\begin{document}


\maketitle

\leftline{\small Mathematical Institute SANU, Serbian Academy of
Sciences and Arts} \leftline{\small Kneza Mihaila 36, 11000
Belgrade, Serbia} \leftline{\small E-mail: bozaj@mi.sanu.ac.rs}

\

\leftline{\small Faculty of Sciences, University of Banja Luka}
\leftline{\small Mladena Stojanovi\'ca 2, 51000 Banja Luka, Bosnia
and Herzegovina}\leftline{\small E-mail: vlajov@blic.net}

\begin{abstract}
In this article we construct L--A representations of geodesic
flows on quadrics and of billiard problems within ellipsoids in
the pseudo--Euclidean spaces.  A geometric interpretation of the
integrability analogous to the classical Chasles theorem for
symmetric ellipsoids is given. We also consider a generalization
of the billiard within arbitrary quadric allowing virtual billiard
reflections.
\end{abstract}

\section{Introduction}

A pseudo--Euclidean space $E^{k,l}$ of signature $(k,l)$,
$k,l\in\N,\, k+l=n$, is the space $\R^{n}$ endowed with the scalar
product
\[
\langle x,y\rangle = \sum_{i=1}^k x_iy_i - \sum_{i=k+1}^{n}
x_iy_i\quad (x,y\in\R^{n}).
\]

Two vectors $x,y$ are {\it orthogonal}, if $\langle x,y\rangle=0$.
A vector $x\in E^{k,l}$ is called \emph{space--like},
\emph{time--like}, \emph{light -- like}, if $\langle x,x\rangle$
is positive, negative, or $x$ is orthogonal to itself,
respectively. Denote by $(\cdot,\cdot)$ the Euclidean inner
product in $\R^{n}$ and let
$$
E=\diag(\tau_1,\dots,\tau_n)=\mbox{diag}(1,\dots,1,-1,\dots,-1),
$$
where $k$ diagonal elements are equal to 1 and $l$ to $-1$. Then
$\langle x,y\rangle=(Ex,y)$, for all $x,y\in\R^{n}$.

Let $M$ be a smooth hypersurface in $E^{k,l}$. A {\it normal}
$\nu(x)$ at $x\in M$ is a vector orthogonal to the tangent plane
$T_x M$. In particular, a normal to the hyperplane  $(n,x)=0$ is
$En$. We say that $x\in M$  is \emph{singular point}, if $\nu(x)$
is light--like, or equivalently, if the induced metric is
degenerate at $x$.

Let $A=\mbox{diag}(a_1,\dots,a_{n})$, $a_i\ne 0, \,i=1,\dots,n$.

Following Khesin and Tabachnikov \cite{KT} and Dragovi\'c and
Radnovi\'c \cite{DR} we consider the geodesic flow and the
billiard system (in the case when $A$ is positive definite) on a
$n-1$--dimensional quadric
%
%
\begin{equation}\label{elipsoid}
\mathbb{E}^{n-1}=\left\{x\in E^{k,l}\,|\, (A^{-1}x,x)=1 \right\}.
\end{equation}
%
Notice that $E\A x$ is a normal at $x\in\mathbb{E}^{n-1}$.
Therefore, $x\in\mathbb{E}^{n-1}$ is singular, if
$(EA^{-2}x,x)=0$.

Lax representations for geodesic lines and billiard
trajectories outside of singular points are constructed
(Theorems 2.1 and 3.1). For billiards, in a general non--symmetric
case, the spectral curve is a non--singular hyperelliptic curve
$\mathcal S$ of genus $n-1$ for a space--like or time--like
trajectory, while for a light--like trajectory its genus is $n-2$.
The billiard mapping transforms to a translation on the Jacobian
variety of $\mathcal S$ by a constant vector (Theorem 3.2).

There is a nice geometric manifestation of the integrability.
Consider the following "pseudo--confocal" family of quadrics in
$E^{k,l}$
\begin{equation}\label{confocal}
\mathcal Q_\lambda: \quad  ((A-\lambda E)^{-1} x,
x)=\sum_{i=1}^n\frac{x_i^2}{a_i-\tau_i\lambda}=1, \quad \lambda
\ne \tau_i a_i, \quad i=1,\dots,n.
\end{equation}
For a non--symmetric ellipsoid, the lines $l_k$, $k\in\mathbb Z$
determined by a generic space--like or time--like (respectively
light--like) billiard trajectory are tangent to $n-1$
(respectively $n-2$) fixed quadrics from the pseudo--confocal
family \eqref{confocal} ({\it pseudo--Euclidean version of the
Chasles theorem}, see Theorem 4.9 in \cite{KT} and Theorem 5.1 in
\cite{DR}). Also, tangent lines to a generic space--like or
time--like (respectively light--like) geodesic are tangent to
other $n-2$ (respectively $n-3$) fixed quadrics from the
pseudo--confocal family \eqref{confocal}. A related geometric
structure of the set of singular points for the pencil
\eqref{confocal} is described in \cite{DR}.

Here we consider the case of symmetric quadrics, when the systems
are integrable in a noncommutative sense (Theorem 4.1) and prove
the Chasles theorem for symmetric ellipsoids (Theorem \ref{ch1}).
By combining Theorem \ref{ch1} and a non-commutative version of
Veselov's discrete Arnold-Liouville theorem (see \cite{Ves3}), we
formulate {\it Poncelet theorem for a symmetric elliptic billiard}
in the pseudo--Euclidean space $E^{k,l}$ (Theorem \ref{poncelet}).

Finally, in the last section, we define a natural generalization
of the billiard within arbitrary quadric allowing the so called
virtual reflections. The virtual billiard flow shows the same
dynamical characteristics as the usual one: the Lax
representation, integrability, and the Chasles theorem (Theorems
6.1, 6.2, and 6.3).

\section{Geodesic flows}
\label{chap:geodezijske}

By the use of the scalar product we can identify tangent and
cotangent spaces $ y\in T_x\R^n \longmapsto p=Ey\in T_x^*\R^n. $
The canonical symplectic form $dp\wedge dx$ on $T^*\R^n(x,p)$
transforms to the form
\begin{equation}\label{symplectic}
\sum_{i=1}^ndp_i\wedge dx_i=\sum_{i=1}^k dy_i\wedge
dx_i-\sum_{i=k+1}^n dy_i\wedge dx_i.
\end{equation}
on $T\R^n(x,y)$.
%
%
It induces the Poisson bracket
\begin{equation}\label{canonical}
\{f,g\}=\sum_{i=1}^k\frac{\p f}{\p x_i}\frac{\p g}{\p
y_i}-\sum_{i=k+1}^n\frac{\p f}{\p x_i}\frac{\p g}{\p y_i} -
\sum_{i=1}^k\frac{\p f}{\p y_i}\frac{\p g}{\p x_i} +
\sum_{i=k+1}^n\frac{\p f}{\p y_i}\frac{\p g}{\p x_i}.
\end{equation}
%
By a {\it geodesic} on $\mathbb{E}^{n-1}$ we mean a critical
smooth curve $\gamma: t\mapsto x(t)\in \mathbb{E}^{n-1}$  of the
action
$$
S[\gamma]=\int_\gamma L(x,\dot x)dt =\int_\gamma\frac12\langle\dot
x,\dot x\rangle dt.
$$

The Euler--Lagrange equation for the Lagrangian $L$ with the
constraint $x(t)\in\mathbb E^{n-1}$ yields
\begin{equation}\label{geodesic}
E\ddot{x}=\mu A^{-1}x,
\end{equation}
where the Lagrange multiplier is $\mu=-(A^{-1}\dot x,\dot
x)/(EA^{-2}x,x)$, provided that $x(t)$ is not a singular point.

By introducing the variable $\dot x=y$, the system
\eqref{geodesic} takes the form
%
%
\begin{equation}\label{hamilton}
\dot x=y,\quad \dot y=\mu
EA^{-1}x=-\frac{(A^{-1}y,y)}{(EA^{-2}x,x)}\, EA^{-1}x,
\end{equation}
%
on the tangent bundle $T\mathbb{E}^{n-1}\setminus \Sigma$
described by the constraints
\begin{equation}\label{constraint}
F_1=(A^{-1}x,x)-1=0,\quad F_2=(A^{-1}x,y)=0,
\end{equation}
where
\begin{equation}\label{sing0}
\Sigma=\{(x,y)\in T\R^n\,\vert\,(EA^{-2}x,x)=0\}.
\end{equation}

The system \eqref{hamilton} is actually a Hamiltonian system with
the Hamiltonian function
%
\begin{equation}
H=\frac12\langle y,y\rangle \label{Ham}
\end{equation}
%
%
with respect to the Poisson--Dirac bracket
\begin{equation}\label{PD}
\{f_1,f_2\}_D=\{f_1,f_2\}-\frac{\{F_1,f_1\}\{F_2,f_2\}-\{F_2,f_1\}\{F_1,f_2\}}{\{F_1,F_2\}},
\end{equation}
where $\{\cdot,\cdot\}$ is the bracket \eqref{canonical} (cf.
\cite{Moser}). Note that
$$
\{F_1,F_2\}=2(EA^{-2}x,x),
$$
and that the system \eqref{hamilton} as well as the bracket
\eqref{PD}, is well defined on $T\R^{n}\setminus\Sigma$. The
functions $F_1$ and $F_2$ are Casimir functions of the
Poisson--Dirac bracket considered on $T\R^n\setminus\Sigma$.


For arbitrary $\lambda\in\R$ let
%
\begin{equation}\label{bilin}
q_{\lambda}(x,y)=((\lambda E-A)^{-1}x,y)=\sum_{i=1}^k\frac{x_iy_i}{\lambda-a_i}-\sum_{i=k+1}^n\frac{x_iy_i}{\lambda+a_i}.
\end{equation}

Similarly as in \cite{BozaRos}, we get
%
\begin{thm}\label{Lax za geod}
Solutions of \eqref{hamilton} on $T\mathbb E^{n-1}\setminus\Sigma$
 satisfy the matrix equation
\begin{equation}
\dot{\mathcal L}_{x,y}(\lambda)=[{\mathcal
L_{x,y}}(\lambda),{\mathcal A_{x,y}}(\lambda)],
\end{equation}
where the $2\times2$ matrices ${\mathcal L_{x,y}}(\lambda),{\mathcal A_{x,y}}(\lambda)$ are given by\mbox{}\\
\[
\begin{array}{rcl}
{\mathcal L_{x,y}}(\lambda)&=&\left(\begin{array}{cc}
q_{\lambda}(x,y) & q_{\lambda}(y,y)\\
-1-q_{\lambda}(x,x) & -q_{\lambda}(x,y)
\end{array}\right),\\[3ex]
{\mathcal A_{x,y}}(\lambda)& = & \left(\begin{array}{cc}
0& {\mu}/{\lambda}\\
1 & 0
\end{array}\right), \quad \mu=-{(A^{-1}y,y)}/{(EA^{-2}x,x)}.
\end{array}
\]
\end{thm}

\begin{cor}\label{integrali} The determinant $\det\mathcal L_{x,y}(\lambda)$
is an integral of the geodesic flow \eqref{hamilton} for all
$\lambda$.
\end{cor}

%
%

We shall say that $\mathbb{E}^{n-1}$ is {\it non-symmetric}, if
$\tau_i a_i\neq \tau_j a_j$ for $i\neq j$. Assuming that
$\mathbb{E}^{n-1}$ is {non-symmetric}, { the matrix representation
described in Theorem \ref{Lax za geod} is equivalent to the system
\eqref{hamilton} up to the discrete group generated by the
reflections
\begin{equation}\label{group}
(x_i,y_i)\,\longmapsto\, (-x_i,-y_i), \qquad i=1,\dots,n.
\end{equation}

Further, }from the expression
\begin{equation}\label{nonSymL}
\det \mathcal
L_{x,y}(\lambda)=q_\lambda(y,y)(1+q_\lambda(x,x))-q_\lambda(x,y)^2=\sum_{i=1}^n\frac{f_i(x,y)}{\lambda-\tau_ia_i},
\end{equation}
one can derive the integrals $f_i$ of the system \eqref{hamilton}
{in the} form
\begin{equation}\label{intgeo}
f_i(x,y)=\tau_i y_i^2+\sum_{j\neq i}\frac{(x_iy_j-x_jy_i)^2}{\tau_j a_i-\tau_i a_j}\quad (i=1,\dots,n).
\end{equation}
It is easy to check that they commute in the Poisson bracket
\eqref{PD}, providing Liouville integrability of the geodesic
flow.
If $A$ is positive definite, $\mathbb E^{n-1}$ is an ellipsoid and
the above integrals coincides with the ones given in \cite{KT}.

It is also convenient to consider a polynomial L--matrix
$$
\mathbb L_{x,y}(\lambda)=\left(\prod_{i=1}^n (\lambda-\tau_i
a_i)\right)\mathcal L_{x,y}(\lambda).
$$
The L--A pair ${\dot{\mathbb L}}_{x,y}=[\mathbb L_{x,y},\mathcal
A_{x,y}]$ belongs to a class of so called Jacobi--Mumford systems
\cite{Mum}. It has a spectral curve
\begin{eqnarray}
&& \mathcal S: \,\,\det(\mathbb L_{x,y}(\lambda)-\eta I)=0
\,\Longleftrightarrow\, \eta^2+\det\mathbb L_{x,y}(\lambda)=0, \label{spectralCurve}\\
&& \det\mathbb L_{x,y}(\lambda)=\left(\prod_{i=1}^n
(\lambda-\tau_i a_i)^2\right)\cdot \det\mathcal
L_{x,y}(\lambda).\label{polynomialP}
\end{eqnarray}

For a non-symmetric quadric, from \eqref{nonSymL}, the polynomial
$\det\mathbb L_{x,y}(\lambda)$ equals
$$
\det\mathbb L_{x,y}(\lambda)=(\lambda-\tau_1 a_1)\cdots
(\lambda-\tau_n
a_n)\cdot(\lambda^{n-1}L_{n-1}+\lambda^{n-2}L_{n-2}+\dots+L_0),
$$
where the integrals $L_i$ depend on $f_i$, in particular $
L_{n-1}=2H=\langle y,y\rangle.$ The integrals $L_i$ are
independent on $T\R^n$, while on $T\mathbb E^{n-1}$, due to
$\det\mathbb L_{x,y}(0)=0$, we have $L_0\equiv 0$.

Therefore, for a space--like or time--like trajectory the degree
of $\det\mathbb L_{x,y}(\lambda)$ is $2n-1$, while for a
light--like trajectory its degree is $2n-2$. For a generic
trajectory all zeros of $\det\mathbb L_{x,y}(\lambda)$ are
different  and $\mathcal S$ is a non--singular hyperelliptic
curve.

\section{Billiards}

Here we suppose that $A$ is positive definite and following
\cite{KT}, consider a billiard flow inside the  ellipsoid
\eqref{elipsoid} in $E^{k,l}$. Between the impacts the motion is
uniform along the straight lines. Suppose also that $x\in{\mathbb
E}^{n-1}$ is non--singular. Then $\nu(x)$ is transverse to the
quadric and the incoming vector $w$ can be decomposed as $w=t+n$,
where $t$ is its tangential and $n$ the normal component in $x$.
The billiard reflection is $w_1=t-n$. If $x\in{\mathbb E}^{n-1}$
is singular, the flow stops.

Let $\phi:(x_j,y_j)\mapsto (x_{j+1},y_{j+1})$ be the billiard
mapping, where $x_j\in{\mathbb E}^{n-1}$ is a sequence of
non--singular impact points and $y_j$ is the corresponding
sequence of outgoing velocities (in the notation we follow
\cite{Veselov, Ves3, fedo}, which slightly differs from the one
given in \cite{MV}, where $y_{j}$ is the incoming velocity).

It is evident from the definition that the Hamiltonian \eqref{Ham}
is an invariant of the mapping $\phi$. Therefore, the lines
$l_k=\{x_k+sy_k\,\vert\,s\in\R\}$ containing segments $x_kx_{k+1}$
of a given billiard trajectory are of the same type: they are all
either space--like ($H>0$), time--like ($H<0$), or light--like
($H=0$).

%
%

As in the Euclidean case (see \cite{Veselov, MV, fedo}), we have:

\begin{lem}\label{pomocna}
{\rm (i)} The billiard mapping $\phi$ is given by:
\begin{eqnarray}
x_{j+1}&=&x_j-2\,\frac{(\A x_j,y_j)}{(\A y_j,y_j)}\, y_j,\label{1bilijar}\\[1ex]
y_{j+1}&=&y_j+2\,\frac{(\A x_{j+1},y_{j+1})}{(EA^{-2}
x_{j+1},x_{j+1})}\, E\A x_{j+1}\label{2bilijar}.
\end{eqnarray}

{\rm (ii)} The function $J_j=(\A x_j,y_j)$ is an invariant of the
billiard mapping.
\end{lem}

\noindent{\it Proof.} (i) Since the normal component of $y_j$ and
$y_{j+1}$ at $x_{j+1}$ is parallel to $EA^{-1}x_{j+1}$, we
conclude that
\begin{equation*}
\begin{array}{rcl}
x_{j+1}-x_j&=& \mu_jy_j,\\[1ex]
y_{j+1}-y_j&=&\nu_jEA^{-1}x_{j+1},
\end{array}
\end{equation*}
for some $\mu_j,\nu_j\in\R$, $j\in\Z$, and the multipliers are
determined from the conditions $(A^{-1}x_{j+1},x_{j+1})=1$ and
$\langle y_j,y_j\rangle=\langle y_{j+1},y_{j+1}\rangle$:
\begin{equation*}\label{mnozioci}
\mu_j=-2\,\frac{(\A x_j,y_j)}{(\A y_j,y_j)}, \qquad
\nu_j=2\,\frac{(\A x_{j+1},y_{j+1})}{(EA^{-2} x_{j+1},x_{j+1})}.
\end{equation*}

(ii) From \eqref{2bilijar} we have
\[
(\A x_{j+1},y_{j+1})=(\A x_{j+1},y_j)+2(\A x_{j+1},y_{j+1}),
\]
hence $(\A x_{j+1},y_{j+1})=-(\A x_{j+1},y_j)$. Further, using
\eqref{1bilijar}, one obtains
\[
(\A x_{j+1},y_{j+1})=-(\A x_{j+1},y_j)=-(\A x_j,y_j)+2(\A
x_j,y_j)=(\A x_j,y_j).
\]
\hfill$\Box$


The initial condition $(x_0,y_0)$ uniquely defines the billiard
trajectory $x_k$. In the other direction, if the initial condition
is given by the two successive non--singular initial points
$x_0,x_1\in\mathbb E^{n-1}$ and $x_1-x_0$ is space--like or
time--like it is natural to take unit length
$y_0=(x_1-x_0)/\sqrt{\vert\langle x_1-x_0,x_1-x_0\rangle\vert}$. If
$x_1-x_0$ is light--like, we simply take $y_0=x_1-x_0$.

Note that in the limit, when $J_j$ tends to zero, the billiard
flow transforms to the geodesic flow on $E^{n-1}$. Conversely,
when the smallest semi--axes of the ellipsoid $\mathbb E^{n-1}$
(say $a_n$) tends to zero, the geodesic flow on $\mathbb E^{n-1}$
transforms to the billiard flow within $(n-2)$--dimensional
ellipsoid $\mathbb E^{n-1} \cap \{x_n=0\}$.

{Motivated by the L--A representation for the Euclidean elliptical
billiard with the Hook potential given by Fedorov \cite{fedo}, we
get: }
\begin{thm}\label{billiardLA}
The trajectories $(x_j,y_j)$ of the billiard map \eqref{1bilijar},
\eqref{2bilijar}, outside the singular set \eqref{sing0}, satisfy
the matrix equation
\begin{equation}\label{billLA}
\mathcal{L}_{x_{j+1},y_{j+1}}(\lambda)=\mathcal{A}_{x_j,y_j}(\lambda)\mathcal{L}_{x_j,y_j}(\lambda)\mathcal{A}_{x_j,y_j}^{-1}(\lambda),
\end{equation}
with $2\times2$ matrices depending on the parameter $\lambda$
\[
\begin{array}{rcl}
\mathcal{L}_{x_j,y_j}(\lambda)&=&\left(\begin{array}{cc}
q_{\lambda}(x_j,y_j) & q_{\lambda}(y_j,y_j)\\
-1-q_{\lambda}(x_j,x_j) & -q_{\lambda}(x_j,y_j)
\end{array}\right),\\[3ex]
\mathcal{A}_{x_j,y_j}(\lambda)&=& \left(\begin{array}{cc}
I_j\lambda+2J_j\nu_j &-I_j\nu_j\\
-2J_j\lambda& I_j\lambda
\end{array}\right),
\end{array}
\]
where $q_{\lambda}$ is given by \eqref{bilin}, and
\[
J_j=(\A x_j,y_j),\quad I_j=-(\A y_j,y_j),\quad
\nu_j=2J_j/(EA^{-2}x_{j+1},x_{j+1}).
\]
\end{thm}

{The theorem can be verified by direct calculations.}

Analogous to the geodesic flow in Section \ref{chap:geodezijske},
from Theorem \ref{billiardLA} we arrive to the integrals
\eqref{intgeo} of the billiard flow \eqref{1bilijar},
\eqref{2bilijar} associated to a non--symmetric ellipsoid
\eqref{elipsoid}.

Symplectic (for space--like and time--like trajectories) and
contact properties (for light--like trajectories) of the mapping
$\phi$ are studied in \cite{KT}. In particular, this is an example
of a contact integrable system \cite{KT2}. Recently, another
integrable discrete contact system, the Heisenberg model in
pseudo--Euclidean spaces, is given in \cite{Jo3}.

By the use of Theorem \ref{billiardLA} we have also an
algebraic--geometrical interpretation of the integrability.

In a non--symmetric case and for generic initial conditions all
zeros of \eqref{polynomialP} are real and different (see
\cite{DR}). Thus, for a space--like or time--like trajectory, the
spectral curve \eqref{spectralCurve} is a hyperelliptic curve of
genus $n-1$, while for a light--like trajectory its genus is
$n-2$.

A generic complexified invariant manifold $L_0=c_0,\dots,
L_{n-1}=c_{n-1}$ of the system factorized by the action of the
discrete group generated by the reflections \eqref{group} is an open
subsets of the Jacobian $J(\mathcal S)$ of the spectral curve
\eqref{spectralCurve} (see \cite{Mum} for the case of the Neumann
system).

Let $E_\pm=(0,\pm\sqrt{-\det\mathbb L(0)})$ and
$$
T=\mathcal A(E_+-E_-),
$$
where $\mathcal A: Div^0(\mathcal S)\to J(\mathcal S)$ is the Abel
mapping.

Repeating the arguments given for Theorem 3 in \cite{fedo}, we
obtain

\begin{thm}\label{translation}
The dynamics \eqref{1bilijar}, \eqref{2bilijar} corresponds to the
translation on the Jacobian variety of the spectral curve
\eqref{spectralCurve} by a vector $T$.
\end{thm}

The Cayley--type conditions for periodic billiard trajectories
within ellipsoids in the pseudo--Euclidean spaces are derived in
\cite{DR}. Theorem \ref{translation} provides an alternative
approach for the derivation of Cayley--type conditions modulo
symmetries \eqref{group} (e.g., see Ch. 3, Section 8 and Ch. 7,
Sections 2 and 3 in \cite{DrRa}).

\section{Symmetric quadrics}
In a more general situation, when the quadric is symmetric, we use
the following notation  (cf. \cite{BozaRos}): the sets of indices
$I_s\subset\{1,\dots,n\}\enspace (s=1,\dots r)$ are defined by the
conditions,
%
%
\begin{equation}\label{sym}
\begin{array}{l}
1^{\circ}\enspace \tau_i a_i=\tau_j a_j=\alpha_s\enspace \mbox{for}\enspace i,j\in I_s\enspace\mbox{and for all}\enspace s\in\{1,\dots,r\},\\[1ex]
2^{\circ}\enspace \alpha_s\neq \alpha_t\enspace \mbox{for}\enspace
s\neq t.
\end{array}
\end{equation}

One should observe the possibility that $a_i=a_j$ for $i\in I_s,\
j\in I_t,\ s\neq t$, but in this case it has to be
$\tau_i\tau_j=-1$.

Owing to Corollary (\ref{integrali}), the determinant
$\mbox{det}\,{\mathcal L_{x,y}}(\lambda)$ is an invariant of the
flow \eqref{hamilton}, and by expanding it in terms of
$1/(\lambda-\alpha_s),1/(\lambda-\alpha_s)^2$, we get
\begin{equation}\label{SymL}
\det\mathcal L_{x,y}(\lambda)=
(1+q_{\lambda}(x,x))q_\lambda(y,y)-q_\lambda(x,y)^2 =\sum_{s=1}^r
\frac{\tilde f_s}{\lambda-\alpha_s}+\frac{P_s}{(\lambda -
\alpha_s)^2},
\end{equation}
where the integrals $\tilde f_s, P_s$ are given by
%
\begin{equation}\label{integrals}
\begin{array}{rcl}
\tilde f_s  &=&\ds\sum_{i\in I_s}\Big(\tau_i y_i^2+\sum_{j\notin I_s}\frac{(x_iy_j-x_jy_i)^2}{\tau_j a_i-\tau_i a_j}\Big),\\[4ex]
P_s  &=&\ds\sum_{i,j\in I_s,i<j}(x_iy_j-x_jy_i)^2\quad
\text{for}\quad |I_s|\ge2 \qquad (P_s\equiv 0, \quad
\text{for}\quad |I_s|=1).
\end{array}
\end{equation}
%

The Hamiltonian \eqref{Ham} is equal to the sum
$H=\frac12\sum_{s=1}^r\tilde f_s$. Also, the functions $\tilde
f_s, P_s$ are independent on $T\R^n$, while restricted to
$T\mathbb{E}^{n-1}$ they are related by
\begin{equation*}
\sum_{s=1}^r \frac{\tilde f_s}{\alpha_s}=\sum_{s=1}^r
\frac{P_s}{\alpha_s^2},
\end{equation*}
which is equivalent to $\det\mathcal L_{x,y}(0)=0$.

An analog of Theorem 5.1 in \cite{BozaRos} holds:
\begin{thm}\label{integraliT}
In addition to \eqref{integrals}, a non--singular geodesic $x(t)$
on a quadric $\mathbb E^{n-1}$  also has integrals
\begin{equation}\label{moment-map}
\Phi_{s,ij}:=y_ix_j-x_iy_j,\qquad i,j\in I_s, \qquad |I_s|\ge2.
\end{equation}

The functions $\tilde f_s$, $P_s=\sum_{i<j}\Phi_{s,ij}^2$ are
central within the algebra of integrals generated by $\tilde f_s$
and $\Phi_{s,ij}$:
\[
\begin{array}{l}
\{\tilde f_s,\tilde f_t\}_D=0,\quad \{\tilde f_s,P_t\}_D=0,\quad \{P_s,P_t\}_D=0,\\
\{\tilde f_s,\Phi_{t,ij}\}_D=0,\quad \{P_s,\Phi_{t,ij}\}_D=0.
\end{array}
\]
\end{thm}
\noindent{\it Proof.} The functions \eqref{moment-map} are
components of the momentum mapping
$$
\Phi_s: \quad TE^{n-1} \longrightarrow so(k_s,l_s)^*
$$
of the Hamiltonian $SO(k_s,l_s)$-action on $TE^{n-1}$, where
\[
k_s=\vert \{\tau_i\,\vert\,\tau_i=1,\,i\in I_s\}\vert, \quad
l_s=\vert \{\tau_i\,\vert\,\tau_i=-1,\,i\in I_s\}\vert, \quad
k_s+l_s=\vert I_s\vert.
\]
Indeed, they are components of the momentum mapping of
$SO(k_s,l_s)$-action on $T\R^{n}(x,y)$ and since the action
preserves the constraints \eqref{constraint}, that is
\begin{equation}\label{mm}
\{ \Phi_{s,ij},F_1\}=\{ \Phi_{s,ij},F_2\}=0,
\end{equation}
they are also components of the momentum mapping of the
Hamiltonian $SO(k_s,l_s)$-action on $T\mathbb E^{n-1}$. In
particular, because $P_s$ is a composition of the momentum mapping
with a Casimir function on $so(k_s,l_s)^*$, we have
$\{P_s,\Phi_{s,ij}\}=\{P_s,\Phi_{s,ij}\}_D=0$.

Since the Hamiltonian function \eqref{Ham}, as well as of all its
components $\tilde f_s$ are invariant with respect to the
$SO(k_s,l_s)$-action, then  the functions \eqref{moment-map} are
integrals of the system and commute with $\tilde f_s$,
$s=1,\dots,r$ (the Noether theorem).

Next, since $\Phi_{s,ij}$ and $\Phi_{t,uv}$ for $s\ne t$ depend on
different sets of variables $(x,y)$, their canonical Poisson
bracket vanishes. Thus, from \eqref{mm} we also have
$\{\Phi_{s,ij},\Phi_{t,uv}\}_D=0$, implying that $\{P_s,P_t\}_D=0,
\{P_s,\Phi_{t,ij}\}_D=0$.

It remains to prove $\{\tilde f_s,\tilde f_t\}_D=0$. Following
\cite{BozaRos}, we introduce a family of deformed non--symmetric
quadrics
\[
{\mathbb E}^{n-1}_{\epsilon}:\,\,\,
(A_{\epsilon}^{-1}x,x)=1,\,\,\,
A_{\epsilon}=\mbox{diag}(a_1^{\epsilon},\dots,a_n^{\epsilon}),\,\,\,
\tau_i a_i^{\epsilon}\ne \tau_j  a_j^{\epsilon},\,\,\, i\neq j
\,\,\,\text{for}\,\,\,\epsilon\ne 0,
\]
where $\lim_{\epsilon\rightarrow 0}a_i^{\epsilon}=a_i$, and
$a_i^{\epsilon}$ are smooth functions. The corresponding
Poisson--Dirac bracket and integrals \eqref{intgeo} are denoted by
$\{\cdot,\cdot\}_D^\epsilon$ and $f_i^{\epsilon}$, respectively.
Define
\begin{equation}
\tilde f^{\epsilon}_s = \ds\sum_{i\in
I_s}f^{\epsilon}_i=\ds\sum_{i\in
I_s}\Big(\tau_iy_i^2+\sum_{j\notin
I_s}\frac{P_{ij}}{\tau_ja^{\epsilon}_i-\tau_ia^{\epsilon}_j}\Big).
\end{equation}
Then $\{\tilde f_s^\epsilon,\tilde f_t^\epsilon\}_D^\epsilon=0$,
and taking the limit $\epsilon\to 0$, we obtain $\{\tilde
f_s,\tilde f_t\}_D=0$.
\hfill$\Box$

\medskip

For a symmetric quadric \eqref{sym}, from \eqref{SymL}, the
polynomial \eqref{polynomialP} determining the spectral curve
\eqref{spectralCurve} equals
$$
\det\mathbb L_{x,y}(\lambda) = (\lambda-\alpha_1)^{2\vert
I_1\vert-\delta_1}\cdots (\lambda-\alpha_r)^{2\vert
I_r\vert-\delta_r}\cdot P(\lambda),
$$
where
\begin{eqnarray}
P(\lambda) &=&
 (\lambda-\alpha_1)^{\delta_1} \cdots
(\lambda-\alpha_r)^{\delta_r}\det\mathcal L_{x,
y}(\lambda)  \label{symP2} \\
 &=& \sum_{s=1}^r
\left((\lambda-\alpha_s)^{\delta_s-1}\prod_{i\ne s}
(\lambda-\alpha_i)^{\delta_i}\tilde f_s+\prod_{i\ne s}
(\lambda-\alpha_i)^{\delta_i}P_s\right) \nonumber \\
&=& \lambda^{N-1}K_{N-1}+\dots+\lambda K_1+K_0,\nonumber
\end{eqnarray}
and
$$
\delta_s=2 \,\,\, \text{for} \,\,\, \vert I_s \vert \ge 2, \quad
\delta_s=1 \,\,\, \text{for} \,\,\, \vert I_s\vert=1, \quad
N=\delta_1+\dots+\delta_r.
$$

In particular, $K_{N-1}=2H=\langle y,y\rangle$. When considered on
$T\R^n$, the functions $K_i$ are independent, while on $T\mathbb
E^{n-1}$, since $P(0)=0$, we have $K_0\equiv 0$.

Thus, the degree of $P(\lambda)$ is $N-1$ for a space--like or
time--like vector $y$, or $N-2$ for a light--like $y$. It can be
proved that the geodesic flow \eqref{hamilton} is integrable in a
noncommutative sense by means of integrals described in Theorem
\ref{integraliT} and that generic invariant isotropic manifolds
are $(N-1)$--dimensional. They are generated by the Hamiltonian
flows of $\tilde f_1,P_1,\dots, \tilde f_r, P_r$, that is, of the
integrals $K_1,\dots,K_{N-1}$.

\section{The Chasles theorem for symmetric ellipsoids}

In this section we assume that $\mathbb E^{n-1}$ is an ellipsoid.
Then the condition $\tau_i a_i=\tau_j a_j$ can be satisfied only
if
\begin{equation}\label{SS}
a_i=a_j, \qquad \tau_i\tau_j=1.
\end{equation}
Therefore, a
symmetric ellipsoid $\mathbb E^{n-1}$ with conditions \eqref{sym}
has $SO(\vert I_1\vert)\times\dots\times SO(\vert
I_r\vert)$--symmetry.

From the discrete L--A representation in Theorem 3.1 we get for
billiards the integrals \eqref{integrals}. Moreover, one can
easily verify that the components \eqref{moment-map} of the
momentum mapping of $SO(\vert I_s\vert)$--action are also
conserved by the billiard flow \eqref{1bilijar}, \eqref{2bilijar},
implying a noncommutative integrability of the mapping $\phi$
booth in the symplectic and in the contact setting (see
\cite{Jov}).

We now give a geometric interpretation of noncommutative
integrability of the systems considered here analogous to the {pseudo--Euclidean
versions of the Chasles theorem} stated in \cite{KT} (see Theorem
4.9) and in \cite{DR} (see Theorem 5.1) for the corresponding Liouville integrable
non--symmetric systems. For the
Euclidean case, see Lemma 6.2 in \cite{BozaRos}.

Consider the pencil of quadrics \eqref{confocal} in $E^{k,l}$. The
condition
\begin{equation}\label{jednacina}
\det \mathcal
L_{x,y}(\lambda)=q_\lambda(y,y)(1+q_\lambda(x,x))-q_\lambda(x,y)^2=0
\end{equation}
is equivalent to the geometrical property that the line
$$
l_{x,y}=\{x+sy, s\in\R\}
$$
is tangent to the quadric $\mathcal Q_{\lambda}$. This is proved
in \cite{Moser, DR} for $\mathbb E^{n-1}$ being a non-symmetric
ellipsoid, but the assertion holds for symmetric quadrics $\mathbb
E^{n-1}$ as well.

\begin{thm}\label{ch1}
{\rm (i)} If a line $l_k$ determined by the billiard segment
$x_{k}x_{k+1}$ (respectively a geodesic line $x(t)$ at the moment
$t=t_0$) is tangent to a quadric $\mathcal Q_{\lambda^*}$ from the
pseudo--confocal family \eqref{confocal}, then it is tangent to
$\mathcal Q_{\lambda^*}$ for all  $k\in\mathbb Z$ (respectively
for all $t\in\R$ ). In addition,  $\mathbf R(x_k)$ is a billiard
trajectory (respectively $\mathbf R(x(t))$  is a geodesic line)
tangent to the same quadric $\mathcal Q_{\lambda^*}$ for all
$\mathbf R\in SO(\vert I_1\vert)\times\dots\times SO(\vert
I_r\vert)$.

{\rm (ii)} The lines $l_k$ determined by a generic space--like or
time--like (respectively light--like) billiard trajectory are
tangent to $N-1$ (respectively $N-2$) fixed quadrics from the
pseudo--confocal family \eqref{confocal}, where, as above
$$
N=r+|\{s\in\{1,\dots,r\}\,:\, \vert I_s\vert\ge 2\}|.
$$
The tangent lines to a generic space--like or time--like
(respectively light--like) geodesic on $\mathbb E^{n-1}$ are
tangent to other $N-2$ (respectively $N-3$) fixed quadrics from
the pseudo--confocal family \eqref{confocal}. Moreover, the
billiard trajectories (geodesic lines) tangent to the same set of
quadrics are of the same type: space--like, time--like or
light--like.
\end{thm}

\noindent{\it Proof.} (i) If the line $l_{x(t_0),y(t_0)}$ is
tangent to $\mathcal Q_{\lambda^*}$ then $\det\mathcal
L_{x(t_0),y(t_0)}(\lambda^*)=0$, implying $\det\mathcal
L_{x(t),y(t)}(\lambda^*)=0$ for all $t\in\R$ (Corollary
\ref{integrali}). Therefore, the line $l_{x(t),y(t)}$ is tangent
to the quadric $\mathcal Q_{\lambda^*}$ for all $t\in\R$.

The second statement follows from the fact that $\det\mathcal
L_{x,y}(\lambda)$ is $SO(\vert I_1\vert)\times\dots\times SO(\vert
I_r\vert)$--invariant function.

(ii) From Lemma \ref{chasles} below it follows that a space--like
or time--like (respectively light--like) line $l_{x(t),y(t)}$
determined by a geodesic line $x(t)$ is tangent to $N-2$
(respectively, $N-3$) fixed quadrics from the pseudo--confocal
family \eqref{confocal} different from $\mathbb E^{n-1}$.

The last statement follows from the distribution of zeros of the
polynomial $P(\lambda)$ described in the proof of Lemma
\ref{chasles}.

The similar assertions hold for billiard trajectories as well.
\hfill$\Box$

\medskip

By combining Theorem \ref{ch1} and a non-commutative version of
Veselov's discrete Arnold-Liouville theorem (see \cite{Ves3}) we
can formulate {\it Poncelet theorem for a symmetric elliptic
billiard} in the pseudo--Euclidean space $E^{k,l}$:

\begin{thm}\label{poncelet}
If  a billiard trajectory $(x_k)$ is periodic with a period $m$ and if
the the lines $l_k$ determined by the segments $x_kx_{k+1}$ are
tangent to $N-1$ quadrics $\mathcal Q_{\lambda_1},\dots,\mathcal
Q_{\lambda_{N-1}}$ (in the space--like or the time--like case) or
to $N-2$ quadrics $\mathcal
Q_{\lambda_1},\dots,\mathcal Q_{\lambda_{N-2}}$ (in the light--like
case), then any other billiard trajectory within $\mathbb E^{n-1}$
with the same caustics is also periodic with the same period $m$.
\end{thm}

%
%
\begin{lem}\label{chasles}
If a point $x$ lies inside, or on the ellipsoid $\mathbb E^{n-1}$,
then the equation \eqref{jednacina} generically has $N-1$
(respectively $N-2$) different real solutions for space--like and
time--like (respectively light--like) vectors $y$. In particular,
if the line $l_{x,y}$ is tangent to $\mathbb E^{n-1}$, then
\eqref{jednacina} generically has $N-2$ (respectively $N-3$)
different real non--zero solutions  for space--like and time--like
(respectively light--like) vector $y$.

\end{lem}
\noindent{\it Proof.}
Here we modify the idea used in \cite{Au, DR} for an analogous
assertion in the case of non--symmetric ellipsoids.

We have
\begin{equation}\label{jed1}
q_\lambda(y,y)=-\ds\sum_{i=1}^n\frac{y_i^2}{a_i-\tau_i\lambda}
=\ds-\sum_{s=1}^r\frac{\langle
y,y\rangle_s}{\alpha_s-\lambda}=\ds-\frac{R(\lambda)}{\ds\prod_{s=1}^r(\alpha_s-\lambda)},
\end{equation}
where
\begin{equation}\label{scalar}
\langle y,y\rangle_s=\sum_{i\in I_s}\tau_i y_i^2
\end{equation}
and
\[
R(\lambda)=\ds\sum_{s=1}^r \langle y,y\rangle_s\prod_{t\neq
s}(\alpha_t-\lambda)= (-1)^{r-1}\cdot\sum_{s=1}^r\langle
y,y\rangle_s\prod_{t\neq s}(\lambda-\alpha_t).
\]

We shall estimate the zeros of $R(\lambda)$. Without losing a
generality, we can assume that for \eqref{sym} we have
\begin{equation}\label{alfe}
\alpha_1>\alpha_2>\dots>\alpha_{\tilde r}> 0>\alpha_{\tilde
r+1}>\dots >\alpha_r.
\end{equation}

From the definition of $R(\lambda)$ we obtain
\[
\mbox{sign}\, R(\alpha_s)=\epsilon_s\,(-1)^{s+r}, \quad
\epsilon_s=\mbox{sign}\,\langle y,y\rangle_s, \quad s=1,\dots,r,
\]
and for a space--like or a time--like vector $y$:
\[
\begin{array}{l}
\mbox{sign}\, R(-\infty)=\mbox{sign}\langle y,y\rangle,\\[1ex]
\mbox{sign}\, R(\infty)=(-1)^{r-1}\,\mbox{sign}\langle y,y\rangle.
\end{array}
\]

Then, since $(x,y)$ is generic, \eqref{SS} and \eqref{alfe}  yield
\begin{equation}\label{epsiloni}
\epsilon_1=\dots=\epsilon_{\tilde r}=+1, \qquad \epsilon_{\tilde
r+1}=\dots=\epsilon_r=-1.
\end{equation}

Therefore, the equation $R(\lambda)=0$ has $r-2$ solutions
$\zeta_s\in(\alpha_{s+1},\alpha_s)$ for
$s\in\{1,\dots,r-1\}\backslash\{\tilde r\}$ and another solution
$\zeta_r\in(-\infty,\alpha_r)$ (if $y$ is space--like) or
$\zeta_0\in(\alpha_1,\infty)$ (if $y$ is time--like).

Firstly, we consider the case when the line $l_{x,y}$ is not
tangent to $\mathbb E^{n-1}$. From the fact that the point $x$
belongs to the interior of the ellipsoid, or to the ellipsoid
itself, it follows that $1+q_0(x,x) \ge 0$. Furthermore, for a
generic $(x,y)$ it is $q_0(y,y)<0$, $q_0(x,y)\ne 0$. Whence,
$$
q_0(y,y)(1+q_0(x,x))-q_0(x,y)^2<0.
$$

By the use of the polynomial \eqref{symP2}, we can rewrite $\det
\mathcal L_{x,y}(\lambda)$ in the form
\begin{equation}\label{qP}
\det \mathcal
L_{x,y}(\lambda)=q_\lambda(y,y)(1+q_\lambda(x,x))-q_\lambda(x,y)^2=\frac{P(\lambda)}{\prod_{s=1}^r(\lambda-\alpha_s)^{\delta_s}}.
\end{equation}

Recall that the degree of $P(\lambda)$ is $N-1$ for a space--like
or a time--like vector $y$, while for a light--like vector $y$ the
degree is $N-2$, $N=\delta_1+\dots+\delta_r$. Thus, for a
space--like or a time--like vector $y$ we have:
\begin{equation}\label{beskonacno}
\det \mathcal L_{x,y}(\lambda) \sim \langle y,y\rangle/\lambda,
\qquad \lambda \to \pm\infty.
\end{equation}

Obviously, the left hand side of \eqref{qP} takes negative values
at the ends of each of the $r-2$ intervals
$$
(\zeta_{r-1},\zeta_{r-2}),\dots,(\zeta_{\tilde r+2},\zeta_{\tilde
r+1}), \,(\zeta_{\tilde r+1},0),\,(0,\zeta_{\tilde
r-1}),\,(\zeta_{\tilde r-1}, \zeta_{\tilde
r-2})\dots,(\zeta_2,\zeta_1),
$$
and in each of the indicated intervals lies
$\alpha_{r-1},\dots,\alpha_2$, respectively.
Since generically $P_s>0$ for $|I_s|\ge 2$, i.e, $\delta_s=2$,
from
\begin{equation}\label{bb}
\lim_{\lambda
\rightarrow\alpha_s-}\frac{\tilde
f_s}{\lambda-\alpha_s}+\frac{P_s}{(\lambda -
\alpha_s)^2}=\infty,\quad \lim_{\lambda
\rightarrow\alpha_s+}\frac{\tilde
f_s}{\lambda-\alpha_s}+\frac{P_s}{(\lambda -
\alpha_s)^2}=\infty,
\end{equation}
and \eqref{SymL} follows that in the interval containing the
corresponding $\alpha_s$ there are at least two zeros of the
polynomial $P(\lambda)$.

In the case $|I_s|=1$, it is $P_s=0$ and the interval contains at least one zero.

The analysis above shows that in $(\zeta_{r-1},\zeta_1)$ there are $\delta_2+\dots+\delta_{r-1}$ zeros. It remains to show that in $(-\infty,\zeta_{r-1})\cup(\zeta_1,\infty)$ lie $\delta_1+\delta_r-2$ (if $y$ is light--like) or $\delta_1+\delta_r-1$ zeros (if $y$ is not light--like).

Indeed, note that when $y$ is not
light--like, it also has negative value at the ends of one
of the intervals $(\zeta_r,\zeta_{r-1})$ (if $y$ is space--like) or
$(\zeta_1,\zeta_0)$ (if $y$ is time--like), containing
$\delta_r$ and $\delta_1$ zeros, respectively (which is in agreement with
\eqref{beskonacno}).

Consequently, in the case $\delta_1=\delta_r=1$ the assertion is clear.

If $\delta_1=2$, $\delta_r=1$ and $y$ is time--like, the conclusion follows from the previous  considerations. On the other hand, if $y$ is light--like or space--like, according
to \eqref{bb}, the additional zero of $P(\lambda)$ lies within the
interval $(\zeta_1,\alpha_1)$.

Similar analysis resolves the cases $\delta_1=1$, $\delta_r=2$ and
$\delta_1=2$, $\delta_r=2$.

Secondly, if $l_{x,y}$ is tangent to $\mathbb E^{n-1}$, then
$q_0(x,y)=0$ and 0 becomes a zero of $P(\lambda)$. The above
analysis concerning the zeros of $P(\lambda)$ remains the same,
except for the interval $(\zeta_{\tilde r+1},\zeta_{\tilde r-1})$.
However, owing to
$$
\frac{dP}{d\lambda}\vert_{\lambda=0}=K_1
$$
(see \eqref{symP2}) and the fact that the integral $K_1$ is generically
different from zero, $P(\lambda)$ changes its sign at $0$.
Therefore, the number of zeros of $P(\lambda)$ lying in the
interval $(\zeta_{\tilde r+1},\zeta_{\tilde r-1})$ is the same as
in the previous case.
\hfill$\Box$

\section{Further generalization: virtual billiards within quadrics}

Note that the billiard mapping \eqref{1bilijar}, \eqref{2bilijar}
is well defined for arbitrary quadric $\mathbb E^{n-1}$ given
by \eqref{elipsoid} and not only for ellipsoids. Hence, segments
$x_{k-1}x_k$ and $x_k x_{k+1}$ determined by 3 successive
points of the mapping \eqref{1bilijar}, \eqref{2bilijar} may be:

\begin{itemize}

\item[(i)] on the same side of the tangent plane $T_{x_k}\mathbb
E^{n-1}$;

\item[(ii)] on the opposite sides of the tangent plane
$T_{x_k}\mathbb E^{n-1}$.

\end{itemize}

In the case (i) we have a part of the usual pseudo--Euclidean
billiard trajectory, while in the case (ii) the billiard
reflection corresponds to the points $x_{k-1} x_k x'_{k-1}$, where
$x'_{k+1}$ is the symmetric image of $x_{k+1}$ with respect to
$x_k$. In the three-dimensional Euclidean case, Darboux referred
to such reflection as a {\it virtual reflection} (e.g., see
\cite{DR2006} and \cite{DrRa}, Ch. 5). In  Euclidean spaces of
arbitrary dimension, such configurations were introduced in
\cite{DR2006}. It appears that a multidimensional variant of
Darboux's 4--periodic virtual trajectory with reflections on two
quadrics, refereed as a double--reflection configuration
\cite{DrRa}, is fundamental in the construction of the double
reflection nets in Euclidean and pseudo-Euclidean spaces (see
\cite{DR2014}). They also played a role in a construction of the
billiard algebra (see Ch. 8, \cite{DrRa}).  The 4--periodic orbits
of real and complex planar billiards with virtual reflections are
also studied in \cite{Gl}.

\begin{dfn}
Let $\mathbb E^{n-1}$ be a quadric in the pseudo--Euclidean space
$E^{k,l}$ defined by \eqref{elipsoid}. We refer to
\eqref{1bilijar}, \eqref{2bilijar} as a {\it virtual billiard
mapping}, and to the sequence of points $x_k$ determined by
\eqref{1bilijar}, \eqref{2bilijar} as a {\it virtual billiard
trajectory} within $\mathbb E^{n-1}$.
\end{dfn}

The virtual billiard dynamics is defined outside the singular set
\begin{equation}\label{singular}
\Sigma=\{(x,y)\in T\R^n\,\,\vert\,\,(EA^{-2}x,x)=0 \,\, \vee
(A^{-1}x,y)=0\,\,\vee\,\,(A^{-1} y,y)=0\}.
\end{equation}

The condition $(A^{-1}y_0,y_0)=0$ implies that the line
$l_0=x_0+sy_0, s\in\R$ does not intersect the quadric $\mathbb
E^{n-1}$ (for example, consider the light--like lines in the space
$E^{1,1}$ and the quadric $x_1^2-x_2^2=1$).

We can interpret \eqref{1bilijar}, \eqref{2bilijar},  in the case
of non light--like billiard trajectories, as the equations of a
discrete dynamical system (see \cite{Veselov, MV, Ves3}) on
$\mathbb E^{n-1}$ described by the discrete action functional:
$$
S[\mathbf x]=\sum_k \mathbf{L}(x_k,x_{k+1}), \qquad \mathbf
L(x_k,x_{k+1})=\sqrt{\vert \langle
x_{k+1}-x_{k},x_{k+1}-x_{k}\rangle \vert },
$$
where $\mathbf x=(x_k), \, k\in\mathbb Z$ is a sequence of points
on $\mathbb E^{n-1}$. Note that a virtual billiard trajectory can
have both virtual and real reflections.

\begin{figure}[ht]
\includegraphics[width=50mm, height=50mm]{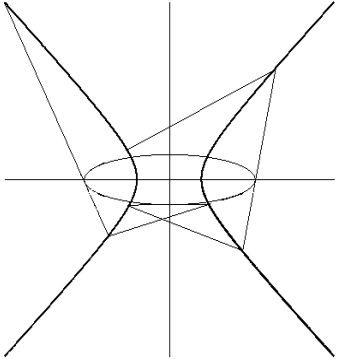}
\caption{A segment of a virtual billiard trajectory within
hyperbola ($a_1>0,a_2<0$) in the Euclidean space $E^{2,0}$. The
caustic is an ellipse. }
\end{figure}

The Lax representation given in Theorem \ref{billiardLA} applies
for the virtual billiard dynamics as well.

\begin{thm}
The trajectories $(x_j,y_j)$ of \eqref{1bilijar},
\eqref{2bilijar}, outside the singular set \eqref{singular}
satisfy the matrix equation \eqref{billLA}.
\end{thm}

Now, suppose that $\mathbb E^{n-1}$ is a symmetric quadric defined
by conditions \eqref{sym}. It has the $G=SO(k_1,l_1)\times
SO(k_2,l_2)\times\dots\times SO(k_r,l_r)$--symmetry (see Theorem
\ref{integraliT}). With the same proof as of the item (i) in
Theorem \ref{ch1}, we have

\begin{thm}
If a line $l_k$ determined by the segment $x_{k}x_{k+1}$ of a
virtual billiard trajectory within $\mathbb E^{n-1}$ (respectively
a geodesic line $x(t)$ at the moment $t=t_0$) is tangent to a
quadric $\mathcal Q_{\lambda^*}$ from the pseudo--confocal family
\eqref{confocal}, then it is tangent to $\mathcal Q_{\lambda^*}$
for all $k\in\mathbb Z$ (respectively for all $t\in\R$). In
addition, $\mathbf R(x_k)$ is a virtual billiard trajectory
(respectively $\mathbf R(x(t))$  is a geodesic line) tangent to
the same quadric $\mathcal Q_{\lambda^*}$ for all $\mathbf R\in
G$.
\end{thm}

However, for a proof of the item (ii), in Lemma \ref{chasles} we used
the relations \eqref{SS}, which implied that, under the conditions
\eqref{alfe}, the signs of the
restricted scalar products \eqref{scalar} satisfy \eqref{epsiloni}.

\begin{figure}[ht]
\includegraphics[width=110mm, height=50mm]{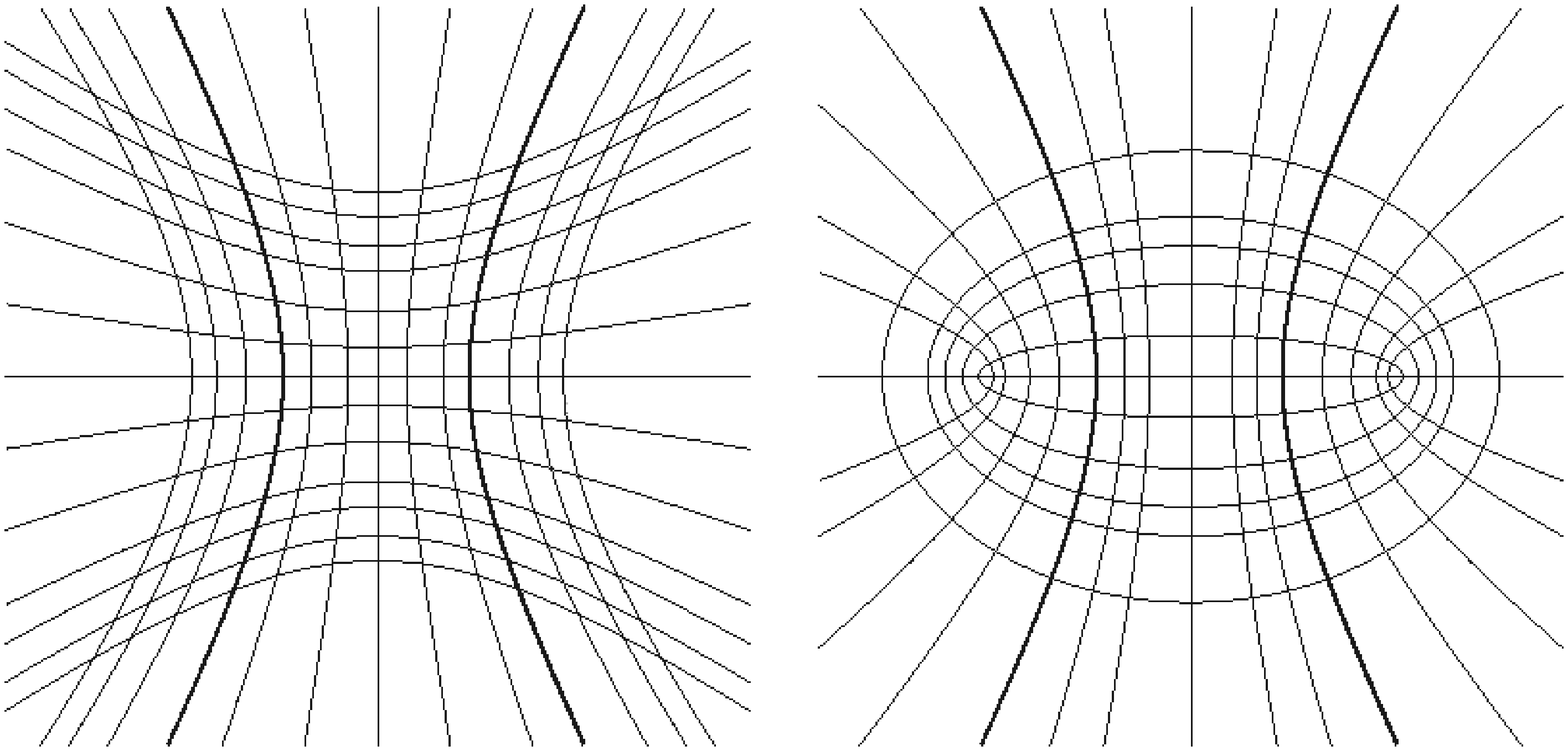}
\caption{Families of pseudo-confocal quadrics for $a_1>0,a_2<0$ in
$E^{1,1}$ and $E^{2,0}$, respectively.}
\end{figure}

For example, let us restrict ourselves to the Euclidean case. With
the same notation as in Lemma \ref{chasles}, for a generic $y$, we
have
$$
\langle y,y\rangle_s > 0, \qquad s=1,\dots,r,
$$
and, therefore,
\begin{eqnarray*}
&& \mbox{sign}\, R(-\infty)=1,\\
&& \mbox{sign}\, R(\alpha_s)=(-1)^{s+r},\\
&& \mbox{sign}\, R(\infty)=(-1)^{r-1}.
\end{eqnarray*}

Consequently, the equation $R(\lambda)=0$ has $r-1$ solutions
$\zeta_s\in(\alpha_{s+1},\alpha_s)$ for $s\in\{1,\dots,r-1\}$ and
the left hand side of \eqref{qP} takes negative values at the ends
of $r-2$ intervals
$$
(\zeta_{r-1},\zeta_{r-2}),\dots,(\zeta_2,\zeta_1).
$$
Also, according to \eqref{beskonacno}, the left hand side of
\eqref{qP} takes negative values at the ends of the interval
$$
(\zeta_r,\zeta_{r-1}),
$$
for a certain $\zeta_r<\alpha_r$. We have $\alpha_s\in
(\zeta_s,\zeta_{s-1})$, $s=2,\dots,r$, and as in the proof of
Lemma \ref{chasles}, this implies that the number of zeros of
$P(\lambda)$ is $N-1$.

Thus, we get:

\begin{thm}
The lines $l_k$ determined by a generic virtual billiard
trajectory within a quadric  $\mathbb E^{n-1}$ in the Euclidean
space $E^{n,0}$ are tangent to $N-1$ fixed quadrics from the
confocal family \eqref{confocal}. Also, the tangent lines to a
generic geodesic on $\mathbb E^{n-1}$ are tangent to other $N-2$
fixed quadrics from the confocal family \eqref{confocal}.
\end{thm}

A sketch of the proof of Theorem 6.2 for a symmetric ellipsoid in
the Euclidean space is given in Lemma 6.2 \cite{BozaRos}.

Finally, we mention that one can obtain similar results for
geodesic flows and billiards on quadrics on a pseudo--sphere in
$E^{k,l}$ (e.g., see \cite{Bo, Ves2, DGJ}). Also, it would be
interesting to describe the class of symmetric periodic (virtual)
billiard trajectories (see \cite{CR} for a study of  symmetric
periodic elliptical billiard trajectories in the Euclidean space).

\subsection*{Acknowledgments}
The research of B. J. was supported by the Serbian Ministry of
Science Project 174020, Geometry and Topology of Manifolds,
Classical Mechanics and Integrable Dynamical Systems.

\end{document}